\documentclass[conference,letterpaper]{IEEEtran}

\usepackage[utf8]{inputenc}
\usepackage[T1]{fontenc}
\usepackage{url}
\usepackage{ifthen}
\usepackage{cite}
\usepackage[cmex10]{amsmath}
\usepackage{algorithmic}
\usepackage{array}
\usepackage[caption=false,font=normalsize,labelfont=sf,textfont=sf]{subfig}
\usepackage{textcomp}
\usepackage{stfloats}
\usepackage{verbatim}
\usepackage{graphicx}
\usepackage{color}
\usepackage{amssymb}
\usepackage{balance}
\begin{document}
\title{A High-Resolution Analysis of Receiver Quantization in Communication}
\author{%
  \IEEEauthorblockN{Jing Zhou}
  \IEEEauthorblockA{Department of Computer Science and Engineering \\
                    Shaoxing University\\
                    Shaoxing, China\\
                    Email: jzhou@usx.edu.cn}
  \and
  \IEEEauthorblockN{Shuqin Pang and Wenyi Zhang}
  \IEEEauthorblockA{Department of Electronic Engineering and Information Science\\
                    University of Science and Technology of China\\
                    Hefei, China\\
                    Email: shuqinpa@mail.ustc.edu.cn, wenyizha@ustc.edu.cn}
}

\maketitle

\begin{abstract}
We investigate performance limits and design of communication in the presence of uniform output quantization with moderate to high resolution.
Under independent and identically distributed (i.i.d.) complex Gaussian codebook and nearest neighbor decoding rule, an achievable rate is derived in an analytical form by the generalized mutual information (GMI).
The gain control before quantization is shown to be increasingly important as the resolution decreases, due to the fact that the loading factor (normalized one-sided quantization range) has increasing impact on performance.
The impact of imperfect gain control in the high-resolution regime is characterized by two asymptotic results:
1) the rate loss due to overload distortion decays exponentially as the loading factor increases,
and
2) the rate loss due to granular distortion decays quadratically as the step size vanishes.
For a $2K$-level uniform quantizer, we prove that the optimal loading factor that maximizes the achievable rate scales like $2\sqrt{\ln (2K)}$ as the resolution increases.
An asymptotically tight estimate of the optimal loading factor is further given, which is also highly accurate for finite resolutions.
\end{abstract}

\section{Introduction}

The analog-to-digital conversion (ADC) is essential for any digital receiver.
In recent years the impact of ADC resolution on the performance has received much attention due to the recent challenges faced in the evolution of wireless communications, including
the increasing processing speed, the increasing scale of hardware, and the need for low cost energy-efficient devices in some new scenarios.
A majority of the studies on ADC at communication receivers have focused on low-resolution quantization, and one-bit quantization has been of particular interest due to its negligible power dissipation and simplicity of implementation, even without requiring automatic gain control (AGC).
In such studies the end-to-end channel is highly nonlinear, typically incurring a substantial performance loss, and necessitating a rethinking of the transceiver design.

On the other side, the transceiver architecture used in present wireless systems is built without considering the effect of output quantization.
It is thus necessary to ask, under such transceiver architecture, how much is the loss caused by output quantization with \emph{moderate to high} resolution?
In other words, if a small loss in achievable rate is acceptable, how fine need the quantization be?
Analytical results on these problems appear to be lacking.
Moreover, limited resolution of quantization leads to new problems, e.g., sensitivity of performance to imperfect gain control, residual interference in multiuser systems, and so on.
These largely unexplored problems prompt us to revisit the topic of receiver quantization in communication.

The performance and design of wireless systems with output quantization have been extensively studied in recent years; see, e.g. \cite{Mezghani-Nossek12,Fan15,Roth-Nossek17,Jacobsson17,Dutta,Chaaban}.
The most common approach therein, however, is not information-theoretic.
Instead, achievable rate estimation based on the additive quantization noise model (AQNM) has been widely used,
which comes from Bussgang-like decomposition \cite{SPM-Bussgang}.
Results in \cite{Mezghani-Nossek12,MNS20,mo17} suggested that such estimation approximates the mutual information well at low signal-to-noise ratio (SNR), but is inaccurate at high SNR.

Although mutual information is a fundamental performance measure, for communications under transceiver nonlinearity it has limited operational meaning, in the sense that the decoder that achieves the predicted rate can be too complex to implement, while that rate is not necessarily achievable by a standard transceiver architecture designed without considering nonlinearity (because the decoder is typically \emph{mismatched} to the nonlinear channel).
In \cite{Zhang12}, a more meaningful performance measure that takes decoding rule into account, namely the generalized mutual information (GMI) \cite{ITFMD}, has been adopted, yielding analytical expressions of achievable rate under output quantization and \emph{nearest neighbor decoding rule}.
Under a given (possibly mismatched) decoder, the GMI determines the highest rate below which the average probability of error, averaged over a given i.i.d. codebook ensemble, converges to zero as the code length $N$ grows without bound \cite{ITFMD}.
In fact, the rate estimation based on the AQNM (or Bussgang-like decomposition) is consistent with the GMI for scalar channel under Gaussian input and nearest neighbor decoding \cite{Zhang12}; see \cite{GNND} for more discussions.
However, such estimation is not accurate in general; see, e.g.,  \cite{LLZ}.

A high-resolution asymptotic theory has already been well established in source quantization \cite{Gray-Neuhoff98}.
A basic result known in \cite{OPS48,Bennett48} states that, for a high-resolution uniform quantizer with step size $\ell$, the mean square error (MSE) can be approximated by $\mathsf{MSE}\approx\ell^2/12$.
This yields the ``$6$-dB-per-bit rule'' that each additional bit in resolution reduces the MSE by $6.02$ dB.
The rule reflects the effect of step size that causes \emph{granular distortion}; but it ignores \emph{overload} distortion due to finite quantization range.
The interplay between these two types of distortion determines how the optimal quantization range scales with the resolution.
The scaling law has been characterized in \cite{Hui-Neuhoff01} for several types of input densities.
In particular, for the Gaussian source it has been shown that 1) for $2K$-level uniform quantization the optimal \emph {loading factor} (normalized one-sided quantization range \cite{GG}) scales like $2\sqrt{\ln (2K)}$ (cf. the conventional ``four-sigma'' rule of thumb \cite{Bennett48,GG}), and
2) with the optimal loading factor, the overload distortion is asymptotically negligible (i.e., $\mathsf{MSE}/(\ell^2/12)\to 1$ in the high-resolution regime).

In this paper, we consider a standard communication transceiver architecture, which includes 1) independent and identically distributed (i.i.d.) complex Gaussian codebook at the transmitter, and 2) a uniform quantizer cascaded with a nearest neighbor decoder at the receiver, where the loading factor of the quantizer can be adjusted by gain control; see Sec. II for details.
Our study is intended to evaluate the loss in achievable rate due to quantization, and characterize the optimal loading factor that maximizes the achievable rate.
In Sec. III we provide an achievable rate expression in analytical form for arbitrary symmetric output quantization, and give asymptotic results as well as numerical evaluations.
The GMI is employed as a basic tool for achievable rate analysis therein.
Our results clearly show the increasing importance of gain control as the resolution decreases.
In Sec. IV-A, the impacts of limited loading factor and finite step size on achievable rate are analyzed in the high-resolution regime.
Sec. IV-B shows that the optimal loading factor that maximizes the achievable rate also scales like $2\sqrt{\ln (2K)}$.
We further provide an asymptotically tight estimate of the optimal loading factor, which is shown to be highly accurate for finite resolutions.
The paper is concluded in Sec. V.

\emph{Notation}:
We use $\phi(t)$ to denote $\frac{1}{\sqrt{2\pi}}\mathrm {exp}\frac{-t^2}{2}$, the probability density function of the standard normal distribution, and use $Q(t)$ to denote the Q-function, i.e., $Q(t):=\int_t^\infty\phi(u)\mathrm d u$.

\section{System Model}
We consider the achievable rate of the ensemble of i.i.d. complex Gaussian codebook with uniform quantization and nearest neighbor decoder at the receiver.
For a code rate $R$ bits/c.u., a message is selected uniformly randomly from the index set $\mathcal M=\{1,2,...,\lceil 2^{NR}\rceil\}$.
If a message $m$ is selected, then the encoder maps it to a length-$N$ codeword $\mathbf X(m)=[X_1(m),...,X_N(m)]$, which is generated according to a product complex Gaussian distribution $\mathcal {N}_\mathbb C(0,\sigma_x^2\mathbf I_N)$.
Each transmitted symbol is scaled by a channel gain $h\in\mathbb C$ which remains constant over the transmission duration of a codeword.
The scaled symbols are corrupted by i.i.d. complex Gaussian noise at the receiver front-end before quantization.
Then the output of the quantizer is
\begin{align}
\label{YqXZ}
Y_n=q\left(g\cdot V_n^{\rm R}\right)+{\textrm j}\cdot q\left(g\cdot V_n^{\rm I}\right),
\end{align}
where $n=1,...,N$,
$V_n^{\rm R}=\mathrm {Re}(hX_n(m)+Z_n)$, $V_n^{\rm I}=\mathrm {Im}(hX_n(m)+Z_n)$,
$q(\cdot)$ denotes the quantizer, $g\in\mathbb R^+$ is a gain-control factor, and the noise $Z_n\sim \mathcal {N}_\mathbb C(0,\sigma^2)$ is independent of $X_n$.
In this model, we let the real and imaginary parts of the received signal be quantized by the same rule with the same gain-control factor.
When the impact of quantization is omitted, the channel capacity is $C=\log(1+\mathsf{SNR})$, where $\mathsf{SNR}=|h|^2\sigma_x^2/\sigma^2$ is the SNR at the receiver front-end.
\subsubsection{Nearest Neighbor Decoder}

The decoder selects a message according to the (scaled) nearest neighbor decoding rule \cite{Lapidoth96,Lapidoth-Shamai02} as
\begin{align}
\label{NND}
\hat{m}=\arg \min\limits_{m\in\mathcal M}\sum\limits_{n=1}^N |Y_n-aX_{n}(m)|^2.
\end{align}
That is, it selects a message corresponding to the codeword (scaled by a parameter $a$) with the minimum Euclidean distance to the received vector $\mathbf Y=[Y_1,...,Y_N]$.

\subsubsection{Symmetric Quantizer and Uniform Quantizer}\label{II2}

Let the quantizer in (\ref{YqXZ}) be symmetric with $2K$ representation points (levels) $\{\pm y_1,...,\pm y_K\}$ and normalized thresholds $\{0, \pm l_1,...,\pm l_{K-1}\}$.
Then its resolution (bit-width) is $b= \log_2{2K}$ bits, which typically satisfies $b\in\mathbb Z^+$.
Let $V\in\mathbb R$ be the input to be quantized and $\sigma_v$ is its standard deviation.
For both quantizers in (\ref{YqXZ}) we have $\sigma_v =\sqrt{(|h|^2\sigma_x^2+\sigma^2)/2}$.
Then for the input $V$, the output of the quantizer is
\begin{align}\label{qs}
q(g V)=y_k\cdot\mathrm{sgn}(V),\; \textrm{if}\;l_{k-1}\sigma_v\leq g|V|<l_k\sigma_v,
\end{align}
where the thresholds satisfy $l_0=0<l_1<...<l_{K-1}<l_K=\infty$.
Apparently, the quantizer output is a nonlinear function of its input, and the thus introduced nonlinearity degrades performance.
In the presence of gain control, we may turn our attention to an equivalent quantization rule for a normalized input $V/\sigma_v$ with adjustable thresholds $\{\ell_k=l_k/g,\; k=1,...,K-1\}$,
where $g$ can be adjusted to optimize the performance.
A special case is the uniform quantizer, which has uniformly located thresholds
\begin{align}
\ell_k&=k\ell, \; k=1,...,K-1,
\end{align}
and mid-rise representation points
\begin{align}
y_k&=\left(k-\frac{1}{2}\right)\ell,\; k=1,...,K,
\end{align}
where $\ell$ is the step size.
Thus, we define its quantization range or support as $[-K\ell,K\ell]$.
Then the loading factor or support limit of the quantizer is $L=K\ell$.

\section{Achievable Rate: Exact and Asymptotic Results}
The following result provides a general approach for achievable rate analysis by the GMI (see \cite{ITFMD} for its formal definition and derivation) in the presence of transceiver distortion with known transition probability.

\textbf{Proposition 1} \cite{Zhang12}:
\emph{For a memoryless SISO channel $X\to Y$ with transition probability $p_{Y|X}(y|x)$ and nearest neighbor decoding rule (\ref{NND}), where $X,Y\in\mathbb C$ and $\mathrm{Var}(X)=\sigma_x^2$, the maximum GMI under i.i.d. complex Gaussian codebook is}
\begin{align}\label{ZhangGMI}
I_\text{GMI}
=\log\frac{1}{1-\Delta},
\end{align}
\emph{where}
\begin{align}
\Delta=\frac{ |\mathrm E[XY^*]|^2}{\sigma_x^2\mathrm E[|Y|^2]}.
\end{align}
\emph{To achieve the maximum GMI given in (\ref{ZhangGMI}), the scaling factor in (\ref{NND}) should be set as}
\begin{align}\label{scaling}
a=\alpha:=\frac{\mathrm E[X^*Y]}{\sigma_x^2}.
\end{align}

Proposition 1 was proved in [\ref{Zhang12}, Appendix C] by direct evaluation and optimization of a general expression of the GMI under the nearest neighbor decoding rule (\ref{NND}).
Based on Proposition 1, we establish the following result which provides an analytical expression for the achievable rate of the transceiver architecture we considered.

\textbf{Theorem 1}:
\emph{For the channel (\ref{YqXZ}) where $q(\cdot)$ is the symmetric quantizer described in Sec. \ref{II2}, the achievable rate under i.i.d. complex Gaussian codebook and nearest neighbor decoding rule (\ref{NND}) is}
\begin{align}\label{main}
I_\text{GMI}=\log(1+\mathsf{SNR})-\log(1+\gamma\mathsf{SNR}),
\end{align}
\emph{where $\gamma$ is a parameter determined by the quantizer as
\begin{align}
\gamma=1-\frac{\mathcal A^2}{\mathcal B},
\end{align}
in which
\vspace{-.1cm}
\begin{align}\label{As}
\mathcal A=\sqrt{2\pi}\sum\limits_{k=1}^{K}y_k\left(\phi(\ell_{k-1})-\phi(\ell_k)\right),
\end{align}
and
\vspace{-.1cm}
\begin{align}\label{Bs}
\mathcal B=\pi \sum\limits_{k=1}^{K}y_k^2\left(Q(\ell_{k-1})-Q(\ell_k)\right).
\end{align}
In particular, if $q(\cdot)$ is the uniform quantizer described in Sec. \ref{II2}, then
\vspace{-.1cm}
\begin{align}\label{Au}
\mathcal A=\sqrt{2\pi}\sum\limits_{k=0}^{K-1}\ell\cdot\phi(k\ell)-\frac{\ell}{2},
\end{align}
and
\vspace{-.1cm}
\begin{align}\label{Bu}
\mathcal B=\pi\sum\limits_{k=0}^{K-1}2k\ell^2Q(k\ell)+\frac{1}{8}\pi\ell^2.
\end{align}
}

\emph{Proof Sketch}: Applying Proposition 1 to the channel (\ref{YqXZ}), by lengthy but straightforward derivation it can be shown that $\mathrm E[XY^*]=\frac{2h^*\sigma_x^2}{\sqrt{\pi(|h|^2\sigma_x^2+\sigma^2)}}\mathcal A$, $\mathrm E[|Y|^2]=\frac{4}{\pi}\mathcal B$, which yield the expression (\ref{main}). $\hfill\blacksquare$

In [\ref{Zhang12}, Sec. V], a parallel result for \emph{real-valued} channel with symmetric output quantization was given.
Theorem 1 shows that the GMI in the complex-valued case has exactly the same expression except that the pre-log factor is doubled.
From Theorem 1, we have the following corollary.

\textbf{Corollary 1}:
\emph{The achievable rate} $I_\text{GMI}$ \emph{given in Theorem 1 has the following properties.}
\begin{itemize}
\item
\emph{The parameter $\gamma$ satisfies
\begin{align}\label{gamma}
0<\gamma\leq 1-\frac{2}{\pi},
\end{align}
where the equality holds when $q(\cdot)$ is a one-bit quantizer, corresponding to an achievable rate}
\vspace{-.1cm}
\begin{align}\label{GMI1bit}
I_\textrm{GMI}^\textrm{1-bit}=\log\frac{1+\mathsf{SNR}}{1+\frac{\pi-2}{\pi}\mathsf{SNR}},
\end{align}
\emph{which converges to $\log_2\frac{\pi}{\pi-2}=1.4604$ bits/c.u. at high SNR.\footnote{The channel capacity converges to $2$ bits/c.u., which is achieved by antipodal signaling. The gap between the GMI and the channel capacity is due to the restricted transceiver architecture.}
As the quantization becomes increasingly fine, we have $\gamma\to 0$ (from above) and}
\vspace{-.1cm}
\begin{align}\label{lossa}
C-I_\text{GMI}=\mathsf{SNR}\cdot\gamma-\mathsf{SNR}\frac{\gamma^2}{2}+o(\gamma^2)\; \;\textrm {nat/c.u.}
\end{align}
\item \emph{High- and low-SNR asymptotics: As $\mathsf {SNR}\to \infty$, we have}
\vspace{-.1cm}
\begin{align}\label{saturate}
I_\text{GMI}=\log\frac{1}{\gamma}-\left(\frac{1}{\gamma}-1\right)\frac{1}{\mathsf{SNR}}+o\left(\frac{1}{\mathsf{SNR}}\right).
\end{align}
\emph{So the saturation rate is} $\overline{I}_\text{GMI}=\log\frac{1}{\gamma}$.
\emph{As} $\mathsf {SNR}\to 0$,
\vspace{-.1cm}
\begin{align}
I_\textrm{GMI}=(1-\gamma) \mathsf {SNR}-\frac{1-\gamma^2}{2}\mathsf{SNR}^2+o\left(\mathsf{SNR}^2\right).
\end{align}
\end{itemize}

\begin{figure}
\centering
\includegraphics[scale=0.43]{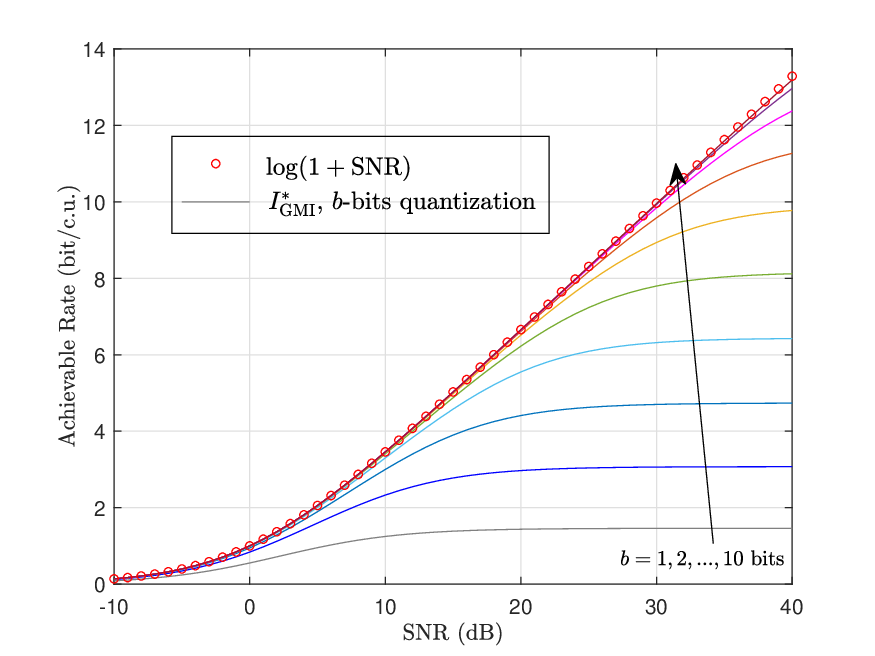}
\caption{Achievable rate (\ref{main}) with $b$-bits uniform output quantization and the optimal loading factor.}
\end{figure}

\begin{figure*}
\begin{minipage}[t]{0.33\linewidth}
$\mspace{-40mu}$\centering
\includegraphics[width=2.28in,height=1.71in]{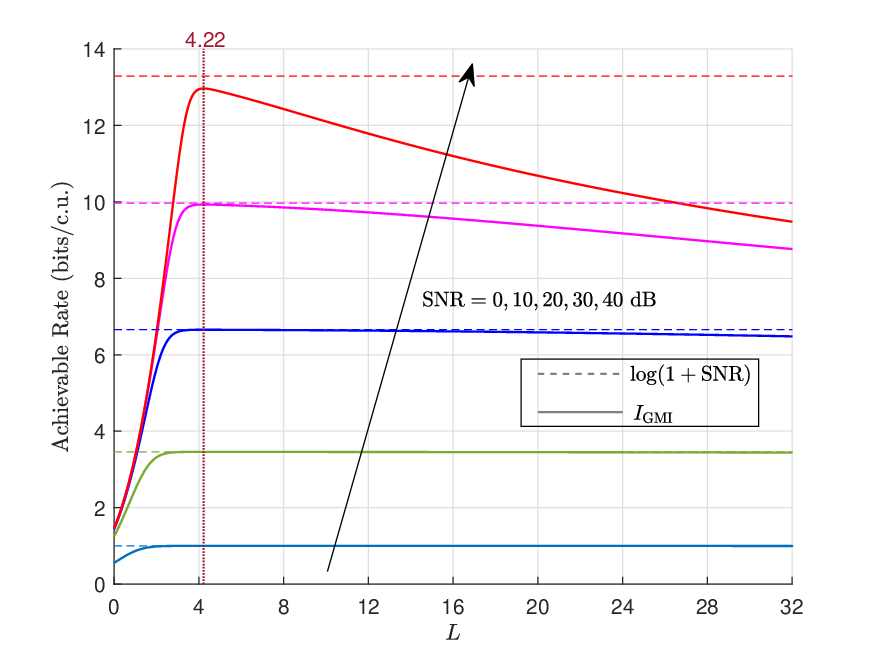}
\caption{Impact of loading factor: $b=9$ bits.$\mspace{40mu}$ Four-sigma rule of thumb is convenient.}
\end{minipage}
\begin{minipage}[t]{0.33\linewidth}
$\mspace{-40mu}$\centering
\includegraphics[width=2.28in,height=1.71in]{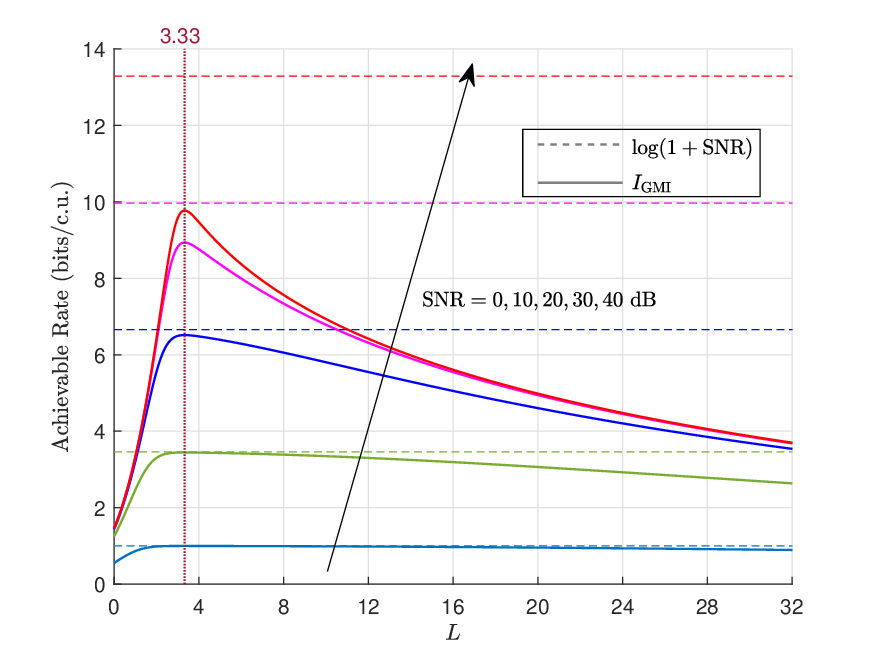}
\caption{Impact of loading factor: $b=6$ bits.$\mspace{40mu}$  Four-sigma rule of thumb causes some loss.}
\end{minipage}
\begin{minipage}[t]{0.33\linewidth}
$\mspace{-40mu}$\centering
\includegraphics[width=2.28in,height=1.71in]{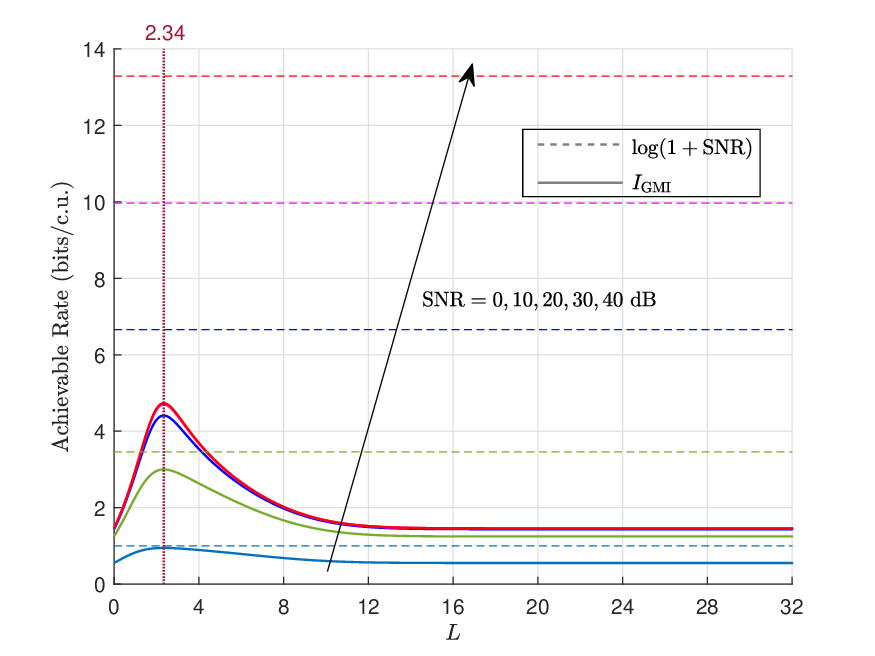}
\caption{Impact of loading factor: $b=3$ bits.$\mspace{40mu}$  Four-sigma rule of thumb causes considerable loss.}
\end{minipage}
\end{figure*}

The parameter $\gamma$ indicates the level of nonlinearity and dominates the asymptotic behavior of performance.
Clearly, it does not depend on the SNR.
Specifically, for a uniform quantizer with a given resolution, (\ref{Au}) and (\ref{Bu}) show that $\gamma$ is determined solely by the step size $\ell$ or equivalently by the loading factor $L$, which can be optimized by adjusting the gain-control factor $g$ in (\ref{YqXZ}) according to the channel gain $h$.

In Figure 1 we show numerical evaluations of $I_\textrm{GMI}$, which have been maximized over $L>0$ for each resolution and is denoted by $I_\textrm{GMI}^*$.
In Figures 2-4, we show how the achievable rate $I_\textrm{GMI}$ varies with the loading factor $L$.
The results clearly reveal the increasing importance of gain control (realized by an AGC module in practical systems) as the resolution decreases:
1) Under high-resolution output quantization, we only require a rough estimate of the channel gain to guarantee that the loading factor to be \emph{no less than} a predefined threshold, say $4$ (from the four-sigma rule of thumb \cite{Bennett48,GG});
2) Such strategy may cause considerable rate loss under low-resolution output quantization, but perfect gain control needs accurate channel estimation, which is challenging in this case.

\section{High-Resolution Asymptotic Results}
In view of the importance of gain control (or equivalently, choice of the loading factor), this section provides results from high-resolution analysis.
We first give the asymptotic decay rates of loss in achievable rate due to limited loading factor and finite step size, respectively.
These results help in understanding the effect of imperfect gain control.
Then the optimal loading factor is characterized, which is essential for reducing the loss due to imperfect gain control.

\subsection{Decay of Rate Loss}

This subsection investigates how the loss in achievable rate scales with loading factor or step size, in the high-resolution regime where $K \to \infty$. The following result shows that the rate loss caused by overload distortion decays exponentially with the loading factor.

\textbf{Theorem 2}:
\emph{In the channel (\ref{YqXZ}) under i.i.d. complex Gaussian codebook and nearest neighbor decoding rule (\ref{NND}), the rate loss due to a $2K$-level uniform quantization with a loading factor $L$ satisfies}
\vspace{-.2cm}
\begin{align}\label{Rlossasy}
C-I_\text{GMI}=\Theta\left(\frac{e^{-\frac{L^2}{2}}}{L^3}\right)\mathsf{SNR} \;\;\textrm {nat/c.u.}%\\
\end{align}
\emph{in the high-resolution limit.}
\begin{IEEEproof}
We note that the summation $\sum_{k=0}^{K-1}\ell\cdot\exp\frac{-k^2\ell^2}{2}$ in $\mathcal A$ is exactly the left Riemann sum of $\exp\frac{-t^2}{2}$ over $[0, K\ell]$ with a regular partition,
and similarly, the summation $\pi\sum_{k=0}^{K-1}2k\ell^2Q(k\ell)$ in $\mathcal B$ is exactly the left Riemann sum of $2\pi tQ(t)$ over $[0, K\ell]$ with a regular partition.
Therefore, for a fixed loading factor $L$ we have the following high-resolution limits:
\vspace{-.2cm}
\begin{subequations}\label{Aa}
\begin{align}
\lim\limits_{K\to\infty} \mathcal A&=\sqrt{2\pi}\lim\limits_ {K\to\infty}\left(\sum\limits_{k=0}^{K-1}\frac{L}{K}\phi\left(\frac{kL}{K}\right)-\frac{L}{2K}\right)\\
&=\sqrt{2\pi}\left(\frac{1}{2}-Q(L)\right),\label{Aa1}
\end{align}
\end{subequations}
\vspace{-.2cm}
\begin{subequations}\label{Ba}
\begin{align}
\mspace{-8mu}\lim\limits_{K\to\infty} \mathcal B
&=2\pi\lim\limits_{K\to\infty}\left(\sum\limits_{k=0}^{K-1}\frac{L}{K}\frac{kL}{K}Q\left(\frac{kL}{K}\right)+\frac{L^2}{16K^2}\right)\\
&=\frac{\pi}{2}-2\pi\int_L^\infty tQ(t)\mathrm d t.\label{Ba1}
\end{align}
\end{subequations}
Combining (\ref{Aa1}) and (\ref{Ba1}), we obtain the high-resolution limit of $\gamma$ as a function of $L$ as
\begin{align}\label{barL}
\bar{\gamma}(L):=\lim\limits_{K\to\infty}\gamma
=\frac{\int_L^\infty\left(\phi(t)-tQ(t)\right)\mathrm d t-Q^2(L)}{\frac{1}{4}-\int_L^\infty tQ(t)\mathrm d t}.
\end{align}
Note that the Q function satisfies \cite{Borjesson-Sundberg79}
\begin{align}\label{BS}
\frac{t}{t^2+1}\phi(t)<Q(t)<\frac{2}{t+\sqrt{t^2+8/\pi}}\phi(t)<\frac{\phi(t)}{t},
\end{align}
implying that its tail behaves like $Q(t)=\frac{1}{t}\phi(t)+\Theta\left(\frac{e^{-t^2/2}}{t^3}\right)$.
Based on this asymptotic behavior, we can simplify (\ref{barL}) for large $L$ and obtain
\begin{subequations}\label{barL1}
\begin{align}
\bar{\gamma}(L)&=4\int_L^\infty \left(\phi(t)-tQ(t)\right)\mathrm d u + O\left(\frac{e^{-L^2}}{L^2}\right)\\
&=\Theta\left(\frac{e^{-\frac{L^2}{2}}}{L^3}\right).\label{asyL}
\end{align}
\end{subequations}
The proof is completed by combining (\ref{barL1}) and (\ref{lossa}).
\end{IEEEproof}

On the other hand, the rate loss caused by granular distortion decays quadratically with the step size.
To prove this we need the following lemma which is one of the various forms of the Euler-Maclaurin summation formula \cite{NA}.

\textbf{Lemma 1}:
\emph{For a real-valued continuously differentiable function} $f(t)$ \emph{defined on} $[a,b]$\emph{, we have}
\begin{align}
\int_a^b f(t) \mathrm d t
=\;& \ell\left(\frac{f(a)}{2}+\sum\limits_{k=1}^{K-1} f(a+k\ell)+\frac{f(b)}{2}\right)\notag\\
&-\frac{\ell^2}{12}\left(f'(b)-f'(a)\right)+ o(\ell^2),
\end{align}
\emph{where} $\ell=\frac{b-a}{K}$.

\textbf{Theorem 3}:
\emph{In the channel (\ref{YqXZ}) under i.i.d. complex Gaussian codebook and nearest neighbor decoding rule (\ref{NND}), the rate loss due to a $2K$-level uniform quantization with a step size $\ell$ satisfies
}
\begin{align}
C-I_\text{GMI}
=\left(\frac{\ell^2}{12}+o(\ell^2)\right)\mathsf{SNR} \;\;\textrm {nat/c.u.}
\end{align}
\emph{in the high-resolution limit.}

\begin{IEEEproof}
Applying Lemma 1, it can be shown that
\begin{align}\label{A}
\sqrt{2\pi}\left(\frac{1}{2}-Q(L)\right)=\mathcal A+o(\ell^2),
\end{align}
\begin{align}\label{B}
&\frac{1}{2}-\int_L^{\infty}2tQ(t)\mathrm d t=\frac{\mathcal B}{\pi}-\frac{\ell^2}{24}+o(\ell^2).
\end{align}
Letting $K\to\infty$, then $L=K\ell\to\infty$ and the above two identities imply that
\begin{align}\label{BAB}
\gamma=\frac{\ell^2}{12}+o(\ell^2).
\end{align}
Substituting it into (\ref{lossa}) completes the proof.
\end{IEEEproof}

\begin{figure}
\centering
\includegraphics[scale=0.45]{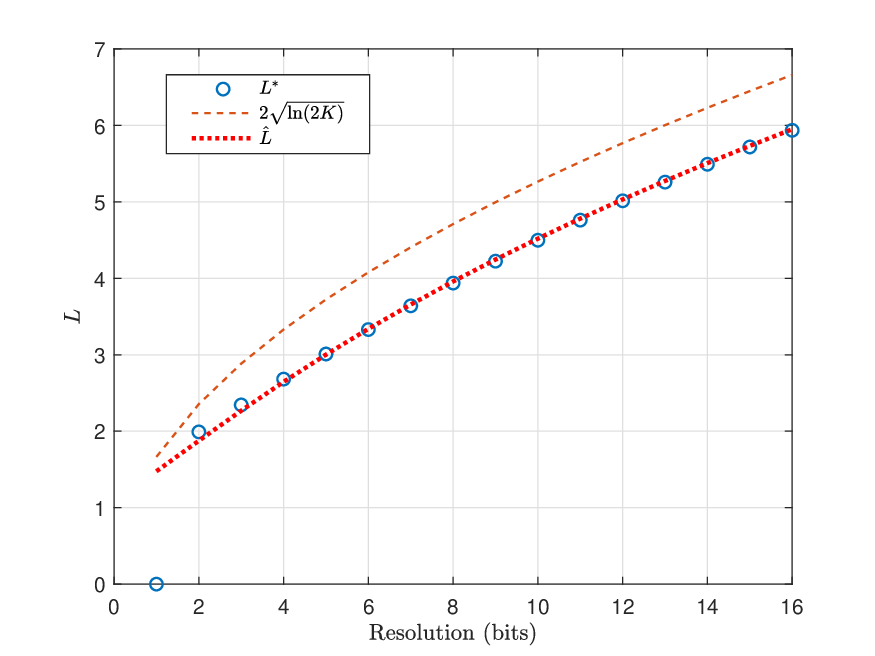}
\caption{Numerical results and asymptotics of $L^\ast$.}
\end{figure}

\subsection{Optimal Loading Factor: Asymptotics and Estimation}

\textbf{Theorem 4}:
\emph{In the channel (\ref{YqXZ}) under i.i.d. complex Gaussian codebook and nearest neighbor decoding rule (\ref{NND}), the optimal loading factor $L^*$ that maximizes the GMI (\ref{main}) satisfies}
\begin{align}
\label{asy}
\lim\limits_{K\to\infty}\frac{L^\ast}{2\sqrt{\ln (2K)}}=1.
\end{align}

\begin{IEEEproof}
We denote the optimal step size by $\ell^*$, i.e., $\ell^*=L^*/K$.
First, from Theorem 3 and Theorem 4, we can infer that (cf. similar results in source quantization theory [\ref{Hui-Neuhoff01}, Example 1])
\begin{align}
\lim\limits_{K\to\infty}K\ell^*=\infty, \;\lim\limits_{K\to\infty}\ell^*=0,\label{llimit}
\end{align}
which enable high-resolution asymptotic analyses.
From Theorem 1, we can prove that
\begin{align}\label{condition}
\frac{\mathrm d I_\text{GMI}}{\mathrm d \ell}=0 \Leftrightarrow \frac{\mathcal A}{\mathcal B}=\sqrt {\frac{2}{\pi}},
\end{align}
which must be satisfied when $L=L^*$.
In the high-resolution regime, by combining (\ref{A}), (\ref{B}), and
 (\ref{condition}) we obtain
\begin{align}
\label{fraclimit}
\frac{\int_{K\ell^\ast}^\infty\exp \frac{-t^2}{2} \mathrm d t +o({\ell^\ast}^2)}{\int_{K\ell^\ast}^\infty 2tQ(t)\mathrm d t -\frac{1}{24}{\ell^\ast}^2+o({\ell^\ast}^2)}=\sqrt{2\pi},
\end{align}
yielding
\begin{align}
\label{conditionK}
\lim\limits_{K\to\infty}\frac{Q(L^\ast) }{\int_{L^\ast}^\infty 2tQ(t)\mathrm d t -\frac{1}{24K^2}{L^\ast}^2}=1.
\end{align}
By noting that
\begin{align}
\lim\limits_{K\to\infty}\frac{Q(L^\ast)}{\int_{L^\ast}^\infty tQ(t)\mathrm d t} =\lim\limits_{K\to\infty}\frac{-\phi(L^\ast)}{-L^\ast Q(L^\ast)}=1,
\end{align}
where the second equality follows from (\ref{BS}),
we obtain
\begin{align}
\label{elementary}
\lim\limits_{K\to\infty}\frac{24K^2 \exp\frac{{-L^\ast}^2}{2}}{\sqrt{2\pi}{L^\ast}^3}=1.
\end{align}
The proof is completed by noting that (\ref{elementary}) implies (\ref{asy}).
\end{IEEEproof}

Theorem 4 determines how $L^*$ scales with $K$.
The scaling law (\ref{asy}) is consistent with the asymptotic behavior of the loading factor of the uniform quantizer that minimize the MSE of a Gaussian source; see \cite{Hui-Neuhoff01}.
The following result provides a useful estimate of $L^*$.

\textbf{Theorem 5}:
\emph{In the channel (\ref{YqXZ}) under i.i.d. complex Gaussian codebook and nearest neighbor decoding rule (\ref{NND}), the optimal loading factor} $L^\ast$ \emph{that maximizes the GMI (\ref{main}) satisfies}
\begin{align}\label{approx}
\lim\limits_{K\to\infty}\left\{L^\ast-\hat{L} \right\}=0,
\end{align}
\emph{where} $\hat{L}$ \emph{is the unique real-valued solution of the transcendental equation}
\begin{align}\label{app}
{L}^2+6\ln L-\ln\frac{18}{\pi}=4\ln(2K).
\end{align}
\begin{IEEEproof}
According to (\ref{llimit}) and (\ref{fraclimit}), the difference between the numerator and the denominator of (\ref{conditionK}) vanishes as $K$ tends to infinity.
Therefore, $L^\ast$ can be approximated with vanishing error by the solution of
\begin{align}\label{app1}
\int_{L}^\infty 2tQ(t)\mathrm d t -Q(L)=\frac{L^2}{24K^2},
\end{align}
since its LHS and RHS are continuous functions of $L$.
According to (\ref{BS}) and its variation
\begin{align}
\left(1-\frac{1}{1+t^2}\right)\phi(t)<2tQ(t)-\phi(t)<\phi(t),
\end{align}
we note that, instead of (\ref{app1}), we can turn to the solution of
\begin{align}
24K^2\phi(L)=L^3,
\end{align}
which is equivalent to (\ref{app}).
\end{IEEEproof}

In Figure 5, we plot $L^*$, $\hat{L}$ in (\ref{approx}), and $2\sqrt{\ln (2K)}$ per (\ref{asy}).
We observe that $\hat{L}$ is a highly accurate estimate for finite resolutions.

\section{Conclusion}
In this paper we explore the robustness of the considered standard communication transceiver architecture under output quantization.
Focusing on moderate-to-high resolution, we highlight the importance of the choice of loading factor (i.e., gain control) when the effect of quantization becomes non-negligible.
The decay of rate loss due to imperfect gain control, and the optimal loading factor for a given resolution, are characterized in the high-resolution regime.
Our results provide insight into the design of receiver quantization in communication.
More scenarios, such as multiantenna and multiuser systems, will be addressed in future work.

\end{document}